\documentclass[10pt,conference]{IEEEtran}

\usepackage{balance}
\usepackage{booktabs}
\usepackage{tabularx}
\usepackage{subfigure}
\usepackage{enumerate}
\usepackage{amssymb} 
\usepackage{amsthm}
\usepackage{enumitem}
\usepackage{xspace}
\usepackage{tikz}
\usepackage{booktabs}
\usepackage{multirow}
\usepackage{lipsum}
\usepackage[htt]{hyphenat}
\usepackage[framemethod=TikZ]{mdframed}
\usepackage{listings}
\usepackage{color}
\usepackage{xcolor}
\usepackage{amssymb}
\usepackage{pifont}
\usepackage{comment}
\usepackage[breaklinks]{hyperref}
\hypersetup{breaklinks=true}
\usepackage[sort,numbers]{natbib}
\usepackage{capt-of}
\usepackage[anythingbreaks]{breakurl}
\usepackage{url}
\PassOptionsToPackage{hyperref}{bookmarksopen,bookmarksdepth=2,breaklinks=true}
\usepackage{xurl}
\usepackage{enumitem,amssymb}

\newlist{todolist}{itemize}{2}
\setlist[todolist]{label=$\square$}

\newcommand{\cmark}{\ding{51}}
\newcommand{\xmark}{\ding{55}}

\definecolor{dkgreen}{rgb}{0,0.6,0}
\definecolor{gray}{rgb}{0.5,0.5,0.5}
\definecolor{mauve}{rgb}{0.58,0,0.82}
\definecolor{applegreen}{rgb}{0.55, 0.71, 0.0}
\definecolor{amber}{rgb}{1.0, 0.75, 0.0}
\definecolor{firebrick}{rgb}{0.7, 0.13, 0.13}
\definecolor{darkblue}{rgb}{0,0,0.55}

\def\eg{\emph{e.g.}\xspace}
\def\etc{\emph{etc.}\xspace}
\def\ie{\emph{i.e.}\xspace}
\def\etal{\emph{et al.}\xspace}
\def\vs{\emph{vs.}\xspace}

\newcommand{\one}{({\em i}\/)}
\newcommand{\two}{({\em ii}\/)}
\newcommand{\three}{({\em iii}\/)}
\newcommand{\four}{({\em iv}\/)}

\newcommand{\pie}[1]{%
\begin{tikzpicture}
 \draw (0,0) circle (0.75ex);\fill (0.75ex,0) arc (0:(-#1+90):0.75ex) -- (0,0) -- cycle;
 \fill (0.75ex,0) arc (0:(#1-90):0.75ex) -- (0,0) -- cycle;
\end{tikzpicture}%
}

\newcommand\crule[1][black]{\textcolor{#1}{\rule{1mm}{1mm}}}

\newcommand\nrule[1][black]{\textcolor{#1}{\rule{3mm}{2mm}}}
\newcommand{\firebrick}[0]{\nrule[firebrick]}
\newcommand{\darkblue}[0]{\nrule[darkblue]}

\newcommand{\LOne}{\crule[applegreen]{}}
\newcommand{\LTwo}{\crule[amber]{} \crule[amber]{}}
\newcommand{\LThree}{\crule[amber]{} \crule[amber]{} \crule[amber]{}}
\newcommand{\LFour}{\crule[red]{} \crule[red]{} \crule[red]{} \crule[red]{}}
\newcommand{\LFive}{\crule[red]{} \crule[red]{} \crule[red]{} \crule[red]{} \crule[red]{}}

\begin{document}

\title{An Empirical Assessment of Global COVID-19 Contact Tracing Applications}

 \author{
     \IEEEauthorblockN{
         Ruoxi Sun\IEEEauthorrefmark{1},
         Wei Wang\IEEEauthorrefmark{1},
         Minhui Xue\IEEEauthorrefmark{1}, 
         Gareth Tyson\IEEEauthorrefmark{2},
         Seyit Camtepe\IEEEauthorrefmark{3}, and
         Damith C. Ranasinghe\IEEEauthorrefmark{1}}
     \IEEEauthorblockA{\IEEEauthorrefmark{1}The University of Adelaide, Australia}
     \IEEEauthorblockA{\IEEEauthorrefmark{2}Queen Mary University of London, United Kingdom}
     \IEEEauthorblockA{\IEEEauthorrefmark{3}CSIRO-Data61, Australia}
 }

\maketitle
 
\begin{abstract}

The rapid spread of COVID-19 has made manual contact tracing difficult. Thus, various public health authorities have experimented with automatic contact tracing using mobile applications (or ``apps''). These apps, however, have raised security and privacy concerns. In this paper, we propose an automated security and privacy assessment tool---\textsc{COVIDGuardian}---which combines identification and analysis of Personal Identification Information (PII), static program analysis and data flow analysis, to determine security and privacy weaknesses. Furthermore, in light of our findings, we undertake a user study to investigate concerns regarding contact tracing apps.
We hope that \textsc{COVIDGuardian}, and the issues raised through responsible disclosure to vendors, can contribute to the safe deployment of mobile contact tracing.
As part of this, we offer concrete guidelines, and highlight gaps between user requirements and app performance.
\end{abstract}

\section{INTRODUCTION}\label{sec:Introduction}

COVID-19 is now a global pandemic affecting over 200 countries, after its first recorded outbreak in December 2019. 
To counter its spread, numerous measures have been undertaken by public health authorities, \eg, quarantining, lock-downs, curfews, physical distancing, and mandatory use of face masks.
Identifying those who have been in close contact with infected individuals, followed by self-isolation (so called \textit{contact tracing}) has proven particularly effective~\cite{WHO2020operations}. Consequently, contact tracing has emerged as a key tool to mitigate the spread of COVID-19. 

Unfortunately, manual contact tracing, using an army of ``detectives'' has proven challenging for many countries~\cite{ferretti2020quantifying,Chappell2020Coronavirus,Latif2020}. 
Therefore, authorities around the world have sought to develop contact tracing applications (or ``apps'') that can automate the process. A plethora of country-centric contact tracing apps and services are currently deployed. These include the Health Code in China~\cite{NYT_report}, the public COVID-19 website in South Korea~\cite{covid19Korea}, and the mobile contact tracing apps released in Australia~\cite{COVIDSafe}, Germany~\cite{corn_warn}, Israel~\cite{hamagen}, Singapore~\cite{bluetrace2020aprotocol}, and United Kingdom~\cite{nhs}. These apps operate by attempting to record prolonged and close proximity between individuals by using proximity sensing methods, \eg, Bluetooth. The data gathered is then used to notify people who may have come in contact with an infected person. 

Proponents argue that the low cost and scalable nature of contact tracing apps make them an attractive tool for health authorities. However, they have proven controversial due to potential violations of privacy~\cite{leins2020tracking} and the security consequences of the mass-scale installation of (rapidly developed) apps across entire populations. Despite attempts to alleviate these concerns, it is well-known that the anonymization of individuals' information is a challenging problem~\cite{culnane2019misconceptions}. 
To mitigate these concerns, we develop a methodology for assessing the security and privacy weaknesses of COVID-19 contact tracing apps. 

\begin{table}[t]
    \small
    \centering
    \caption{Contact tracing apps considered in our study.}
    \label{table:app_list}
    \vspace{-3mm}
    \resizebox{\linewidth}{!}{
    \begin{tabular}{@{}rllr|rllr@{}}
        \toprule
        \# & \textbf{Applications} & \textbf{Country} & \textbf{Downloads} & \# & \textbf{Applications$^{\ast}$} & \textbf{Country} & \textbf{Downloads}\\ \midrule
        1 & COVIDSafe & Australia & 1M & 
        21 & STOP COVID19 CAT & Spain & 500K \\
        2 & Hamagen & Israel & 1M & 
        22 & CG Covid-19 ePass & India & 500K \\
        3 & TraceTogether & Singapore & 500K & 
        23 & \begin{tabular}[c]{@{}l@{}}StopTheSpread \\COVID-19\end{tabular} & UK & 100K \\
        4 & StopCovid France & France & 1M & 
        24 & Stop COVID-19 KG & Kyrgyzstan & 10K \\
        5 & Next Step (DP3T) & Switzerland & N/A &
        25 & BeAware Bahrain & Bahrain & 100K \\
        6 & Corona Warn App & German & 5M &
        26 & \begin{tabular}[c]{@{}l@{}}Nepal COVID-19 \\Surveillance\end{tabular} & Nepal & 5K \\
        7 & \begin{tabular}[c]{@{}l@{}}NHS Test and \\Tracing App\end{tabular} & UK & 10K &
        27 & Stop Covid & Georgia & 100K \\
        8 & TraceCORONA & Germany & N/A &
        28 & Contact Tracing & USA & 10K \\
        9 & Private Kit & USA & 10K &
        29 & Contact Tracer & USA & 10K \\
        10 & MySejahtera & Malaysia & 500K &
        30 & Coronavirus Algérie & Algeria & 100K \\
        11 & Smittestop & Denmark & 10K &
        31 & CoronaReport & Austria & 10K \\
        12 & Covid Alert & Canada & 500K &
        32 & Covid19! & Czech & 10K \\
        13 & SwissCovid & Switzerland & 500K &
        33 & Coronavirus Bolivia & Bolivia & 50K \\
        14 & Bluezone & Vietnam & 100K &
        34 & Coronavirus - SUS & Brazil & 1M \\
        15 & COCOA & Japan & 5M &
        35 & COVA Punjab & India & 1M \\
        16 & Immuni & Italy & 1M &
        36 & SOS CORONA & Mali & 10K \\
        17 & Stopp Corona & Austria & 100K &
        37 & Hamro Swasthya & Nepal & 50K \\
        18 & Aarogya Setu & India & 100M &
        38 & COVID Radar & Netherlands & 50K \\
        19 & EHTERAZ & Qatar & 1M &
        39 & \begin{tabular}[c]{@{}l@{}}NICD COVID-19 \\Case Investigation\end{tabular} & South Africa & 10K \\
        20 & \begin{tabular}[c]{@{}l@{}}Vietnam Health \\Declaration\end{tabular} & Vietnam & 100K &
        40 & Coronavirus UY & Uruguay & 100K \\
        \bottomrule
    \end{tabular}}
\end{table}

Specifically, this paper consists of the following key tasks: \one~we develop a tool, \textsc{COVIDGuardian}, which uses static and dynamic program analysis, as well as a keyword database utilizing natural language processing (NLP) technology, to identify security weaknesses and personally identifiable information (PII) leakage in apps; 
\two~we conduct a comprehensive security and privacy assessment across 40 state-of-the-practice global contact tracing apps (listed in Table~\ref{table:app_list}); 
\three~based on the assessment results, we conduct a user study involving 373 participants, to investigate user concerns and the requirements of contact tracing apps. Through our study, we aim to answer the following research questions:

\begin{itemize}[leftmargin=*]
    \item \textbf{RQ1:} What is the performance of our security and privacy assessment methodology, \textsc{COVIDGuardian}, compared to state-of-the-practice mobile app assessment tools?
    \item \textbf{RQ2:} What is the security and privacy status of state-of-the-practice contact tracing apps?
    \item \textbf{RQ3:} What is the robustness of state-of-the-practice contact tracing apps against potential security and privacy threats? 
    \item \textbf{RQ4:} What are the user concerns and requirements of contact tracing apps? 
\end{itemize}

\noindent{\textbf{The main contributions of our study are as follows.}}

\begin{itemize}[leftmargin=*]
\item
    We develop \textsc{COVIDGuardian},\footnote{Publicly available at \url{https://covid-guardian.github.io/}.} the \textit{first} automated security and privacy assessment tool that tests contact tracing apps for security weaknesses, malware, embedded trackers and private information leakage. \textsc{COVIDGuardian} outperforms 4 state-of-the-practice industrial and open-source tools. 
\item
    We assess the security and privacy status of 40 worldwide Android contact tracing apps. We discover more than 50\% of the apps pose potential security risks due to: \one~employing cryptographic algorithms that are insecure or not part of best practice (72.5\%); and \two~storing sensitive information in clear text that could be potentially read by attackers (55.0\%). Over 40\% of apps pose security risks through Manifest weaknesses, \eg, allowing permissions for backup (hence, the copying of potentially unencrypted application data). Further, we identify that approximately 75\% of the apps contain at least one tracker, potentially causing privacy violations, \ie, leaks that lead to exposing PII to third parties. 
\item 
    By reviewing the state-of-the-art, we identify four major privacy and security threats against contact tracing apps. Our threat analysis finds that apps adopting decentralized architectures are not necessarily more secure than those adopting centralized architectures (by our measures).
    We also conduct a user study involving 373 participants, to investigate user concerns and requirements. The survey results indicate that the tracing accuracy and potential privacy risks are the two major concerns. Compared to users' expectations of accurate proximity recording, users are more likely to use contact tracing apps with better privacy by design.
    
    \item We have disclosed our security and privacy assessment reports to the related stakeholders on 23 May 2020. We have received acknowledgements from numerous vendors, such as MySejahtera (Malaysia), Contact Tracer (USA), and Private Kit (USA). Our re-assessments shown in Table~\ref{table:recheck} confirm that their updates have addressed several of the issues identified.
    
\end{itemize}

We believe our study can provide useful insights for government policy makers, developers, and researchers to build secure contact tracing apps, and to contain infectious diseases in the present and future. 

\section{Background and Related Work}\label{sec:solutions}

\subsection{Taxonomy of Contact Tracing Apps}

The country-centric contact tracing apps we study fall into two broad categories: \one~centralized and \two~decentralized. 

\noindent \textbf{Centralized architectures.} Many apps utilize a centralized system in which a central server is responsible for: \one~collecting the contact records from diagnosed users; and \two~evaluating the health status of users and selecting who to notify. 
For example, in China and South Korea, centralized contact tracing systems were rapidly developed and released. These systems helped health authorities to successfully control the spread of COVID-19. However, a huge amount of PII was collected~\cite{Washington_posts,NYT_report}. 
 
For instance, \textit{TraceTogether}~\cite{bluetrace2020aprotocol} from Singapore and \textit{COVIDSafe}~\cite{COVIDSafe} from Australia rely on proximity tracing via Bluetooth broadcasts from apps. That said, designs that expose PII may not work well in countries with certain societal norms. Thus, many western countries developed solutions with no PII related information exchange, \eg,~\textit{PEPP-PT}~\cite{PEPP-PT} and the ROBERT system~\cite{ROBERT} implemented by \textit{StopCovid France}. 

\noindent \textbf{Decentralized architectures.} 
The second type of solution is decentralized, where \one~the back-end server is only responsible for collecting the anonymous identifiers of diagnosed users; and \two~the data is processed locally on the device to identify who to alert. 
This design prevents the central server from knowing at-risk persons or their contacts. 

These decentralized apps operate in a range of ways. For example, \textit{Hamagen}~\cite{hamagen} relies on location information. Users download the location history of diagnosed users from a back-end server to check if they been in  contact with any infected people.
Other decentralized solutions, \eg, \textit{DP3T}~\cite{troncoso2020decentralized} and the Exposure Notification framework by Google and Apple (Gapple)~\cite{AppleGoogle}, utilize Bluetooth beacons.
In this design, users periodically download the anonymous identifiers of infected people and compare them against previously encountered beacons to compute the risk of exposure.
Note that \textit{NHS~COVID-19}~\cite{nhs} and \textit{Corona Warn App}~\cite{reelfs2020corona} implement the Gapple framework. This design paradigm reduces the privacy exposure of users as only anonymous identifications are shared.  

\subsection{Related Work}\label{sec:related_work}

\begin{figure*}
	\centering
	\includegraphics[width=0.99\linewidth]{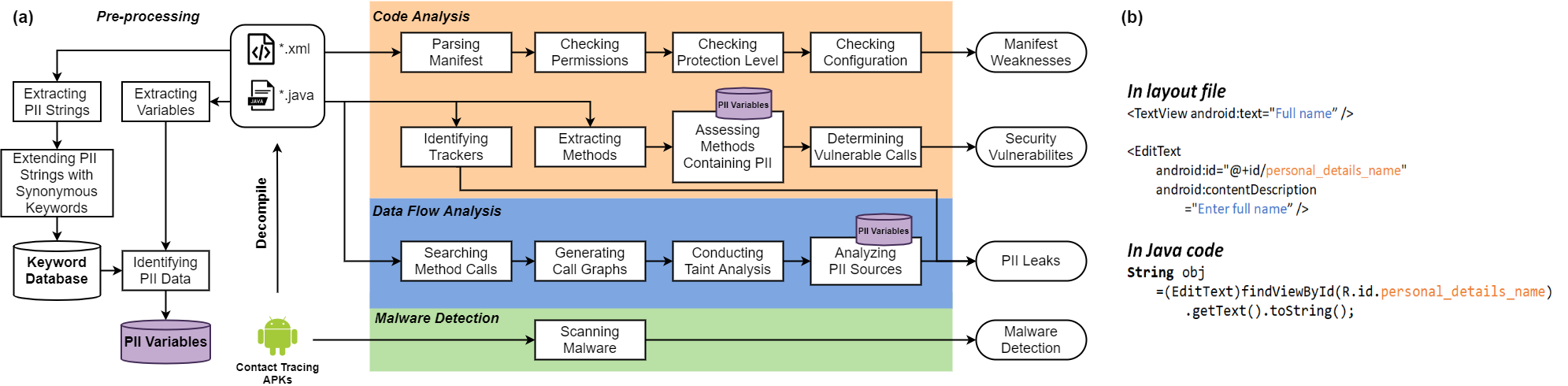}
	\vspace{-2mm}
	\caption{(a)~\textsc{COVIDGuardian}: An overview of our security and privacy assessment methodology; (b)~An example pattern of rendering a view widget.}
	\label{fig:overview}
	\vspace{-4mm}
\end{figure*}

A number of prior works have utilized similar methodologies to \textsc{COVIDGuardian}. We discuss them below. 

\noindent \textbf{Security and privacy analysis for mobile apps.} 
% static analysis and dynamic analysis
\textsc{COVIDGuardian} relies on static code analysis. This is performed by examining source code for signs of security vulnerabilities without executing the program.
In contrast, dynamic analysis executes the code.
Whereas static analysis often suffers from false positives, dynamic analysis is limited by the execution coverage~\cite{reardon201950}. 
Several studies~\cite{chen2020empirical,li2015iccta,xue2016you,chen2018mobile,liu2020maddroid,tang2020ios,wang2020clipboards} have used static analysis to analyze different types of software in search of malicious behaviours and privacy leaks. Static analysis techniques are also widely used in the practical assessment of mobile apps. For example, \textsc{Qark}~\cite{qark} is a static code analysis tool designed to discover various mobile security-related vulnerabilities, either in source code or APKs. 
\textsc{AndroBugs}~\cite{androbugs} is a framework that helps developers find potential mobile security vulnerabilities by pattern-matching. \textsc{MobSF}~\cite{mobsf} offers automated application penetration testing, malware analysis, and a security and privacy assessment framework. \textsc{FlowDroid}~\cite{arzt2014flowdroid} statically computes data flows in apps to understand which parts of the code that data may be exposed to.

Notably, these off-the-shelf tools only utilize syntax-based scanning and data-flows. Therefore, they cannot verify any identified vulnerabilities, which leads to numerous false positives that are not relevant to PII data leakage.
Thus, \textsc{COVIDGuardian} complements this with other methodologies, including the use of third party malware detection and techniques for data flow analysis.

\noindent \textbf{Contact tracing apps analysis.} 
Several works~\cite{gvili2020security,baumgartner2020mind} have focused on security and privacy analysis of the Bluetooth and cryptography specifications published by Apple and Google~\cite{AppleGoogleBluetooth,AppleGoogleCryptography}, arguing that significant risks may be present.
Other work~\cite{li2020covid,liuprivacy} has conducted a review of centralized and decentralized solutions, and proposed privacy-preserving contact tracing using a zero-knowledge protocol. Sun~\etal~\cite{sun2020venuetrace} proposed a privacy-by-design solution, termed \textsc{VenueTrace}, which enables contact tracing of users based on venues they have visited. \textsc{VenueTrace} preserves user's privacy by avoiding information exchanges between users and no private data is exposed to back-end servers.
He \etal~\cite{he2020beyond} inspected the broad implications of COVID-19 related apps, identifying the presence of malware. 
Wang \etal~\cite{wang2020market} also performed a statistical analysis of contact tracing app popularity, as well as user reviews.

\noindent \textbf{Our work.} In contrast to recent works on analyzing contact tracing apps~\cite{zhiqiang,dp3tanalysis,gvili2020security,baumgartner2020mind,cho2020contact,dp3tanalysis}, we not only propose and develop a holistic automated security and privacy assessment tool, \textsc{COVIDGuardian}, but also undertake user studies to perceive users' concerns and requirements to reinforce security and privacy by design. \textsc{COVIDGuardian} works for both centralized and decentralized apps.

\section{Methodology of \textsc{COVIDGuardian}}\label{sec:security_analysis}

In this section, we propose an automated security and privacy assessment tool, \textsc{COVIDGuardian}, to identify security and privacy risks in contact tracing apps. Using \textsc{COVIDGuardian}, we conduct a security and privacy assessment of 40 contact tracing apps from Google Play Store and evaluate their security performance against four categories: \one~manifest weaknesses; \two~general security vulnerabilities; \three~data leaks (with a focus on PII); and \four~malware detection (see Table~\ref{table:assessment_items}). Further, we compare \textsc{COVIDGuardian} with several state-of-the-practice tools to evaluate its ability and performance. 

An overview of \textsc{COVIDGuardian} is shown in Figure~\ref{fig:overview}(a). We first compile a set of PII items, then perform: \one~code analysis to detect Manifest weaknesses and security risks; \two~data flow analysis to reveal the privacy leaks in contact tracing apps; and \three~malware detection.

\vspace{1mm}
\noindent \textbf{PII identification (pre-processing).}
Considering the large-scale use of COVID-19 contact tracing apps and the concerns about PII data collection, we pre-process the contact tracing apps to construct a PII keyword database. 
Figure~\ref{fig:overview}(a) shows the process of PII keyword database construction and the approach to identifying variables that contain PII (defined as \textit{PII Variables}). 

To construct the keyword database, we first focus on the PII input by users through widgets in the user interface, \eg, \texttt{EditText}.
We decompile the Android APKs and extract all strings defined in layout files and resource files, including the widget ID name of any \texttt{EditText} components, the hint text, and the text in the \texttt{TextView}. We manually filter the strings and keep the ones related to PII as seed keywords, \eg, name, phone number, postcode, and password. We then utilize \textsc{Word2vec}~\cite{mikolov2013efficient} to expand our keyword pool with synonymous words.
To ensure the generalization of our PII keyword database, we train the model with the text extracted from 54,371 general apps by the project, \textsc{Supor}~\cite{huang2015supor}. \textsc{Word2vec} presents each word as a vector, and the vectors of synonymous words will have a small cosine distance. 
After training, for each of the seed keywords, we identify the top 5 synonymous words, and then feed them into the keyword database after manual validation. Other numbers of synonyms could be applied in practice, but a larger number will increase the workload of manual review. 
Using Google translate, we check that all apps that have other languages also have contexts (strings) in English, and that all variable names in source code are in English. Considering other languages may increase the efforts of manual filtering, and we therefore only extract English keywords.

After the keyword database is built, we identify which variables in the code are related to PII. 
We do this by keyword matching, which binds semantics to variables. Concretely, we first extract variables related to the widgets in the user interface, \eg, \texttt{EditText} for user input, and \texttt{TextView} for hints or display text. We then determine a variable as a PII Variable if it matches with any keyword in the database. 
For example, in Figure~\ref{fig:overview}(b) we present a typical layout XML file for a widget and the Java code to access it. The text attributes in the layout file can help users quickly understand the usage of a widget, \eg, the string ``Full name'' for an input text box. When the app is compiled, the widget is referenced by an integer ID which is typically assigned in the layout XML file as a string, in the \texttt{id} attribute, \eg, \texttt{personal\_details\_name}. By keyword matching (\eg, ``name'') we tag the variable \texttt{obj} as a PII Variable, as it gets content from the \texttt{EditText} widget where a user will input their full name. Although the tool is tailored to focus on contact tracing apps, our synonym compilation and keyword matching framework can also be adapted to other contexts.

\noindent \textbf{Code analysis.}
We next perform static analysis on the Android Package (APK) binary files. We first decompile the APK of each app to its corresponding \textit{class} and \textit{xml} files. As shown in Figure~\ref{fig:overview}(a), the de-compiled \texttt{AndroidManifest.xml} file is first parsed to extract essential information about the app, such as \texttt{Permission}, \texttt{Components}, or \texttt{Intents}. Then, we assess requested permissions and examine whether all \texttt{Components} (\eg, \texttt{Service}, \texttt{Receiver}, \texttt{Activity}, \texttt{Provider}) are protected by at least one permission explicitly requested in Manifest files. Other attribute configurations, such as the \texttt{allowBackup}, \texttt{debuggable}, and \texttt{networkSecurityConfig} flags, will also be checked.

The Extracting Method module matches methods in decompiled files with pre-defined rules to extract potentially vulnerable methods. For example, if a method contains the keyword \texttt{.hashCode()}, it relies on the Java Hash Code (a weak hash function). The Assessing Methods Containing PII module utilizes the PII Variable database to identify methods containing potential PII. However, as a weakness could be defined in a third-party API, the vulnerable method inspected may never actually be executed during run-time. To address this, the Determining Vulnerable Calls module assesses whether a vulnerable method is actually called and determines whether the PII data is accessed. \textsc{COVIDGuardian} records all the vulnerabilities listed in the \textit{Manifest Weaknesses} and \textit{Vulnerabilities} categories in Table~\ref{table:assessment_items}.

The assessed vulnerabilities include SQL injection, IP address disclosure, hard-coded encryption keys, improper encryption, use of insufficiently random values (CWE~330)~\cite{insecurerandom}, insecure hash functions, and remote WebView debugging being enabled. To increase accuracy, \textsc{COVIDGuardian} not only relies on the detection of a vulnerable method being called, but also employs PII data matching. 
For example, a vulnerable method detection rule may use keywords ``log'' or  ``print'' to locate the method calls related to ``data logging'', \eg, \texttt{Log.v()} and \texttt{System.out.print()}. We further check whether the logged data contains PII by matching the inputs with the PII Variables.

Finally, the trackers in apps (\eg, Google Firebase Analytics, Facebook Analytics, and Microsoft Appcenter Analytics) are detected by the Tracker Identification module and recorded in the \textit{Privacy Leaks} category in Table~\ref{table:assessment_items}. 

\noindent \textbf{Data flow analysis.}
We next conduct a data flow analysis to identify high risk privacy leaks. The data flow analysis extracts the paths from data sources to sinks, and the code statements transmitting the data outside of the app. We define \textit{sources} as calls to any PII data we identify, \eg, \texttt{getViewById(int)} and \texttt{getText()}. Furthermore, we also consider methods that may obtain personal information without user input, \eg, \texttt{getLatitude()} for geographic location information and \texttt{database.Cursor.getString()} for database queries. 
Note that, although unauthorized users cannot directly access sources (only sinks), PII data may still leak during storage or transmission activities. For example, an app may not have permission to access location data directly (the source), but it could obtain that information from the SMS outbox (the sink).
If PII data flows into a code point where unauthorized users or apps can access it (\eg, via local storage, external storage or SMS), the confidentiality of the PII is broken. Here we define \textit{sinks} as methods that may leak sources through specific channels, \eg, \texttt{SharedPreferences\$Editor.putString()}, \texttt{Bundle.putAll()}, and \texttt{SmsManager.sendTextMessage()}.

Thus, we next search the app for lifecycle and callback methods. Using this, we generate a call graph. Starting at the detected sources, the analysis tracks taints by traversing the call graph. If PII data flows from a source to a sink, it indicates that there is a potential privacy leak path. To reduce false-positives, we conduct a backward flow analysis. If the vulnerable code is reachable (\ie, not dead code) and contains PII Variables, we determine it is a potential valid privacy leak. For example, if we find there is a PII Variable that flows into a sink (\eg, Bundle, Log output, SMS) where unauthorized users can access, we will trace it backwards to its source and confirm whether the source is reachable. If reachable, we consider it as a privacy leak.  

\begin{table}[t]
    \tiny
    \centering
    \caption{Security and privacy assessment category.}
    \label{table:assessment_items}
    \vspace{-3mm}
    \resizebox{\linewidth}{!}{
        \begin{tabular}{l|l|l|l|l|l|l}
        \toprule
        \tiny\textbf{\begin{tabular}[c]{@{}l@{}}Assessment \\ Category\end{tabular}} & \tiny\textbf{Security and Privacy Risks} & 
        \rotatebox[origin=c]{90}{\tiny\textbf{ \textsc{COVIDGuardian} }} & 
        \rotatebox[origin=c]{90}{\tiny \textsc{MobSF}} & 
        \rotatebox[origin=c]{90}{\tiny \textsc{AndroBugs}} & 
        \rotatebox[origin=c]{90}{\tiny \textsc{Qark}} & 
        \rotatebox[origin=c]{90}{\tiny \textsc{FlowDroid}} \\ \hline
        \multirow{3}{*}{\textbf{\begin{tabular}[c]{@{}l@{}}Manifest\\ Weaknesses\end{tabular}}} & \begin{tabular}[c]{@{}l@{}}Insecure flag settings \\ (e.g., app data backup allowed)\end{tabular}    &  \pie{360}  & \pie{90} & \pie{90} & \pie{90} &  \\ \cline{2-7} 
        & Non-standard launch mode      &  \pie{360}  & \pie{90} &  & \pie{90} &  \\ \cline{2-7} 
        & Clear text traffic            &  \pie{360}  & \pie{90} & \pie{90} &  &  \\ \hline
        \multirow{8}{*}{\textbf{\begin{tabular}[c]{@{}l@{}}Security\\ Vulnerabilities\end{tabular}}} & Sensitive data logged &  \pie{360}  & \pie{90} &  & \pie{90} &   \\ \cline{2-7} 
        & SQL injection                 &  \pie{360}  & \pie{90} & \pie{90} &  &   \\ \cline{2-7} 
        & IP address disclosure         &  \pie{360}  &  &  &  &   \\ \cline{2-7} 
        & Uses hard-coded encryption key &  \pie{360}  & \pie{90} & \pie{90} & \pie{90} &   \\ \cline{2-7} 
        & Uses improper encryption      &  \pie{360}  & \pie{90} & \pie{90} & \pie{90} &   \\ \cline{2-7} 
        & Uses insecure SecureRandom    &  \pie{360}  & \pie{90} &  &  &   \\ \cline{2-7} 
        & Uses insecure hash function   &  \pie{360}  & \pie{90} & \pie{90} &  &   \\ \cline{2-7} 
        & \begin{tabular}[c]{@{}l@{}}Remote WebView debugging \\ enabled\end{tabular}   &  \pie{360}  & \pie{90} &  & \pie{90} &   \\ \hline
        \multirow{2}{*}{\textbf{PII Leaks}} & Trackers  &  \pie{360}  & \pie{90} &  &  &   \\ \cline{2-7} 
        & \begin{tabular}[c]{@{}l@{}}Potential Leakage Paths from\\ Sources to Sinks\end{tabular} &  \pie{360}  &  &  &   & \pie{90} \\ \hline
        \textbf{\begin{tabular}[c]{@{}l@{}}Malware\\ Detection\end{tabular}}    & \begin{tabular}[c]{@{}l@{}}Viruses, worms, Trojans and \\other kinds of malicious content\end{tabular} &  \pie{360}  &  &  &   &  \\ \bottomrule
    \end{tabular}}
\begin{tabular}{l}
   { \scriptsize{\pie{360} Automated flow analysis, \pie{90} Syntax-based scanning} }
\end{tabular}
\vspace{-3mm}
\end{table}

\noindent \textbf{Malware detection.}
To complement the code analysis, we rely on malware scanners to flag malicious artifacts in contact tracing apps. \textsc{COVIDGuardian} sends the APKs to \textsc{VirusTotal}~\cite{virustotal}, a free online service that integrates over 70 antivirus scanners. Note this has been widely adopted by the research community~\cite{liu2020maddroid,ikram2019chain,hu2019want}. As shown in Table~\ref{table:assessment_items}, the results of malware detection identify viruses, worms, Trojans, and other malicious content embedded in the apps.

\noindent\textbf{Implementation.}
\textsc{COVIDGuardian} includes two components: static code and PII data-flow analysis engines. The static code analysis engine employs \textsc{jadx}~\cite{jadx} to decompile the \texttt{dex} byte code of APK files to Java code. This allows the engine to generate an abstract syntax tree (AST) and create call graphs. Then, the engine scans the given source code to find risks listed in the OWASP~\cite{owasp}. Finally, the taint analysis engine utilizes the call graphs previously generated, with the list of sinks and sources, to locate private data leakage. 

To perform the detection, we first summarize 220 types of (suspected) function calls, which we then use to identify vulnerabilities in the source codes. Then 195 taint sources and sinks are collected to analyze PII leakage through the AST. Finally, the static code analysis engine verifies whether the app logic invokes the data leakage functions or not. 

\section{Evaluation and Results}

\subsection{Selection of apps under-Study}

To evaluate \textsc{COVIDGuardian}, and explore the risks associated with popular contact tracing apps, we curate a list of apps to study.
To achieve this, we first search for keywords in the Google Play Store, \eg, ``contact tracing'', ``Covid'', and ``tracing coronavirus''. 
We also search for known official apps from countries, \eg, the COVIDSafe recommended by the Australian government. After a contact tracing app is found, we assess its functionality by reading the app description and select those with in excess of 10,000 downloads. We also include two beta apps. 
Subsequently, we include the app into the set and look for new apps through the recommendation links in the app store. We repeat this until there are no more contact tracing apps found. 
At last, we finalized the list of 40 contact tracing apps, as shown in Table~\ref{table:app_list}. More detailed information, such as the versions of apps, is provided in our open source website.

\subsection{RQ1: Evaluation of \textsc{COVIDGuardian}} \label{sec:rq1}

\noindent \textbf{Comparison with the state-of-the-practice tools.} 
To evaluate the effectiveness and accuracy of \textsc{COVIDGuardian}, we compare against four industrial and open-source state-of-the-practice security assessment tools. 
These tools are selected based on the number of stars on GitHub (\eg, MobSF 7.6K, Qark 2.4K), their update frequency, usability, and their ability to analyze the security of Android apps. MobSF is recommended by OWASP~\cite{mobsf_owsap}; Qark and AndroBugs are widely used open-source security assessment tools for Android apps; FlowDroid is a classic tool for static taint analysis. We therefore believe they offer an effective baseline to compare against.

We apply all four tools to the 40 contact tracing apps under-study. The detection precision results are listed in Table~\ref{table:tools_comparison_}. The number of \textit{types} is the sum of the number of risks and leaks identified by \textsc{COVIDGuardian}. The precision rates are obtained through manual validation (\ie, filtering out all false positives). 
The worst performing tool is \textsc{FlowDroid}, which generates a large number of false positives that fall  into ``Log'' related sinks.
This is because \textsc{FlowDroid} marks all log methods as sinks without considering whether the logged data is PII or not. For instance, in TraceTogether, error messages such as \texttt{SQLiteException} (from the stack trace) are logged by the \texttt{Log.e} method which matches the keywords ``log'' and is falsely identified as a privacy leak. 

In contrast, \textsc{COVIDGuardian} is able to identify most log related false positives through analyzing the PII Variables (precision of 96.19\%). It is also optimized to filter out non-private data, such as \texttt{Locale.getCountry}. Therefore, most false positives are removed automatically, and only eight false positives (which are from database query results but are not sensitive) are found. 

Another type of false positive found in \textsc{COVIDGuardian} is related to SQL Injections. 
For instance, TraceTogether encapsulates all SQL manipulation methods to limit the input to the SQL query. 
Both \textsc{MobSF} and \textsc{AndroBugs} regard them as at risk of SQL Injections, since they analyze apps by keyword scanning. We manually analyzed all 40 apps and found that ten false positives (of which inputs are limited to several constants in the app but are still regarded as injected) fall into this category in \textsc{COVIDGuardian}.  

\begin{table}[]
    \centering
    \begin{minipage}[t]{0.48\linewidth}
        \caption{Tool Comparison.} 
        \label{table:tools_comparison_}
        \vspace{-2mm}
        \resizebox{\linewidth}{!}{
            \begin{tabular}{@{}lcr@{}}
            \toprule
                \textbf{Tools} & \textbf{\# Types} & \textbf{Precision} \\ \midrule
                \textsc{COVIDGuardian} & 315 & 96.19\% \\
                \textsc{MobSF} & 213 & 47.41\% \\
                \textsc{AndroBugs} & 76 & 80.26\% \\
                \textsc{FlowDroid} & 201 & 40.32\% \\
                \textsc{Qark} & 93 & 84.94\% \\ \bottomrule
            \end{tabular}
        }
        %\vspace{-3mm}
    \end{minipage}\quad
    \begin{minipage}[t]{0.48\linewidth}
        \caption{Identified Trackers.} 
        \label{table:tracker}
        \vspace{-2mm}
        \resizebox{\linewidth}{!}{
            \begin{tabular}{@{}lcr@{}}
                \toprule
                \textbf{Trackers} & \textbf{\# Apps} & \textbf{Percentage} \\ \midrule
                Google Firebase & 25 & 71.4\% \\
                Google CrashLytics & 6 & 17.1\% \\
                Other Google trackers & 4 & 11.4\% \\
                Facebook trackers & 3 & 8.6\% \\
                Other trackers & 9 & 25.7\% \\ \bottomrule
            \end{tabular}
        }
    \end{minipage}
    
\end{table}

\noindent\textbf{Threats to validity.} 
Considering that both the code analysis and data flow analysis rely on keyword matching, a potential cause of false negatives is poorly chosen of keywords. This could mean that some vulnerabilities are not defined in the analysis rules. Similarly, in our data flow analysis, although we update the sources and sinks extracted by SuSi with newly defined ones based on the PII data we identify, there may exist PII leakage that does not match any sources or sinks. We aim to improve the false negatives by updating the rules and keywords database in the future. Currently, our empirical assessment accentuates the identified vulnerabilities and privacy leakage paths.

\begin{mdframed}[backgroundcolor=white!10,rightline=true,leftline=true,topline=true,bottomline=true,roundcorner=2mm,everyline=true] 
	\textbf{Answer to RQ1.} State-of-the-practice tools are less effective (\ie, lower precision) compared to \textsc{COVIDGuardian}. Our assessment methodology can be used to identify security weaknesses with high precision.
\end{mdframed}

\subsection{RQ2: Empirical Assessment Results} 

We next explore the presence of security vulnerabilities among the 40 considered apps using \textsc{COVIDGuardian}.  

\begin{figure}
    \centering
    \includegraphics[width=0.95\linewidth]{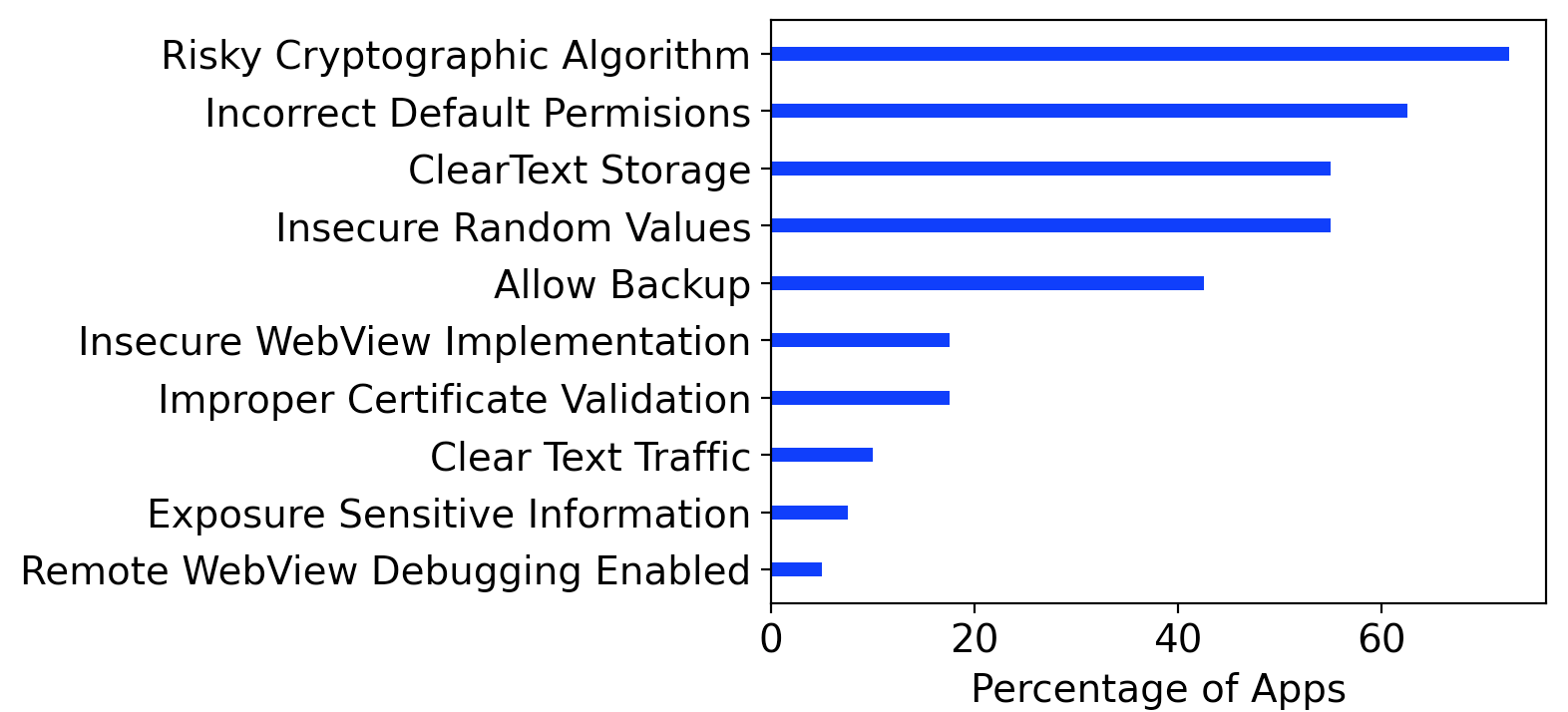}
    \vspace{-3mm}
    \caption{Percentages of apps that are subject to vulnerabilities based on code analysis.}
    \label{fig:analysis_results}
\end{figure}

\noindent \textbf{Code analysis results.} 
Figure~\ref{fig:analysis_results} shows the percentage of contact tracing apps that have security weaknesses found via our code analysis.
We observe that 42.5\% of apps do not set the flag \texttt{allowBackup} to \texttt{False}. Consequently, users with enabled USB debugging can copy application data from the device. Other weaknesses identified are related to ``Clear Text Traffic'', such as plaintext HTTP, FTP stacks, DownloadManager, and MediaPlayer. These may enable a network attacker to implement man-in-the-middle (MITM)~\cite{mitmconcept} attacks during network transmission. 

Figure~\ref{fig:analysis_results} shows that the most frequent weakness identified by code analysis is the ``Risky Cryptography Algorithm''. Over 72.5\% of apps use at least one deprecated cryptographic algorithm, \eg, MD5 and SHA-1. For instance, in the app MySejahtera (Malaysia), the parameters in \texttt{WebSocket} requests are combined and encrypted with MD5. These will be compared with the content from requests in the class \texttt{Draft\_76} in order to confirm the validity of connections. Although this has been listed in the top 10 OWASP~\cite{owasp} mobile risks 2016, the results show that it is still a common security issue. 
Another frequent weakness is ``Clear Text Storage'' (\ie, creation of files that may contain hard-coded sensitive information like usernames, passwords, keys, \etc). For example, in the \texttt{DataBaseSQL} class of the COVID-19 (Vietnam) app, the SQLite database password is stored in the source code without encryption; CG Covid-19 ePass (India) also hard-coded its encryption key in its \texttt{Security} class.

In total, 20 trackers have also been identified, including \texttt{Google Firebase Analytics}, \texttt{Google CrashLytics}, and \texttt{Facebook Analytics}. Approximately 75\% of the apps contain at least one tracker. In the most extreme case, one app, Contact Tracing (USA), contains 8 trackers. 
As shown in Table~\ref{table:tracker}, the most frequent tracker is  \texttt{Google Firebase Analytics}, which is identified in more than 70\% of the apps. Notably, a research study~\cite{leith2020coronavirus} argues that TraceTogether, by  using Google's Firebase service to store user information, maybe leaking user information.

\begin{figure}
    \centering
    \includegraphics[width=0.95\linewidth]{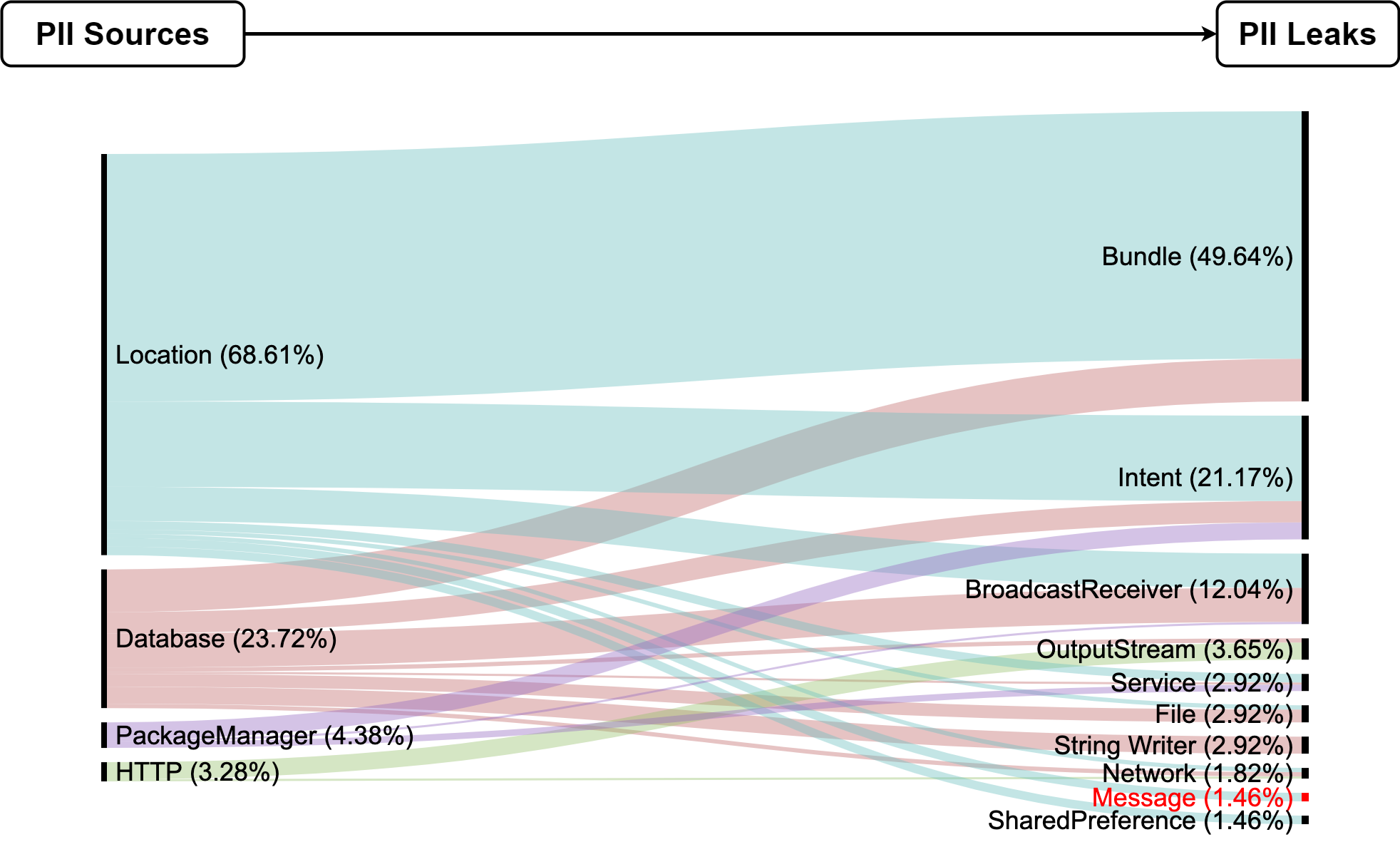}
    \vspace{-2mm}
    \caption{Privacy leaks detected between sources and sinks. Percentages indicate the fraction of flows originating at the sources (left) and terminating at the sinks (right).}
    \label{fig:data_flow}
    \vspace{-3mm}
\end{figure}

\noindent \textbf{Data flow analysis results.} 
Figure~\ref{fig:data_flow} presents the potential privacy leakage between sources and sinks. This is counted by the number of source-to-sink paths found in each app. The top sources of PII data are methods calling from \texttt{Location} and \texttt{database.Cursor}. 
These may obtain PII from a geographic location sensor or from a database query. Most of the PII data will be transferred to sinks, such as \texttt{Bundle}, \texttt{Intent}, and \texttt{BroadcastReceiver}, which may leak PII out of the app. 

As discussed above, sending PII  to the \texttt{Bundle} object may reveal PII data to other activities. 
Notably, we also discover that some apps transmit location information through SMS messages. Considering Hamagen (Israel) as an example, location information is detected and obtained by a source method called \texttt{initialize(Context,Location,e)}. This then flows to a sink method where \texttt{Handler.sendMessage(Message)} is called. This is a potential vulnerability, as malware could easily intercept the outbox of the Android SMS service~\cite{arzt2014flowdroid}.  

\noindent \textbf{Malware detection results.} 
We discover only one app with malware, Stop COVID-19 KG (Kyrgyzstan)~\cite{Stop_COVID-19_KG}. Two risks are identified: a variant Of \texttt{Android/DataCollector.Utilcode.A} and an Adware (0053e0591). This aligns with the finding of COVID-19 apps' threats reported elsewhere~\cite{AnomaliThreatResearch}. 
Considering the limited use of Stop COVID-19 KG (roughly 10K downloads), we conclude that the vast majority of contact tracing apps downloaded from Google Play Store are free of malware. That said, the rise of contact tracking apps has also attracted the interest of malicious developers, \eg, a recent report~\cite{ransomwareCanada} disclosed that new ransomware has targeted the contact tracing app in Canada even before its public release.

\noindent \textbf{Feedback from developers.} After alerting all developers of ours findings, we re-checked the apps by regression testing.
Table~\ref{table:recheck} summarizes the regression testing results.
We find that all potential sources of privacy leakage on three apps --- TraceTogether (Singapore), BlueZone (Vietnam), STOP COVID19 CAT (Spain) --- have been fixed. Additionally, the trackers in MySejahtera (Malaysia) have been removed and the vulnerable app, Contact Tracer (USA), is no longer available.

Meanwhile, new vulnerabilities are identified in the updated versions of several apps (see Table~\ref{table:recheck}). For example, COVA Punjab (India) enables popup windows in the WebView setting. STOP COVID19 CAT (Spain) allows clear text traffic in the Manifest, and some apps have more trackers identified. 
This may mean that the urgency of app development has impacted quality assurance procedures.
We note that a study of 493 iOS apps~\cite{iOS_investigation} confirms that security issues are prevalent in iOS too. Investigating the root cause of the security issue and their urgency is beyond the scope of our research, yet it is an interesting future research topic.

\begin{mdframed}[backgroundcolor=white!10,rightline=true,leftline=true,topline=true,bottomline=true,roundcorner=2mm,everyline=true] 
	\textbf{Answer to RQ2.}  
    The most frequent weaknesses found in contact tracing apps are risky cryptography algorithms use (72.5\%), incorrect default permissions (62.5\%), clear text storage (55.0\%), insecure random values (55\%), and allowing backup (42.5\%). Meanwhile, trackers and potential privacy leakage also pose risks to users' personal information.
\end{mdframed}
 \vspace{-3mm}

\begin{table}[t]
    \centering
    \caption{Results of regression testing of apps.} 
    \label{table:recheck}
    \footnotesize
    \vspace{-2mm}
    \begin{tabular}{@{}lll@{}}
        \toprule
        \textbf{Applications} & \textbf{Version} & \textbf{Issues Patched} \\ \midrule
        
        TraceTogether & 2.0.15 & 
        \begin{tabular}[c]{@{}p{5cm}@{}}
            \cmark~Disabled Allow Backup; \\ 
            \cmark~Fixed potential privacy leakage.
        \end{tabular} \\ \midrule
        
        \begin{tabular}[l]{@{}l@{}}STOP COVID19 \\ CAT
        \end{tabular} & 2.0.3 &
        \begin{tabular}[c]{@{}p{5cm}@{}}
            \cmark~Fixed Insufficient Random issue as they do not use Microsoft package anymore;\\ 
            \cmark~Fixed potential privacy leakage; \\ 
            \xmark~Allows clear text traffic in manifest; \\
            \xmark~New tracker detected: Google CrashLytics.
        \end{tabular} \\ \midrule
        
        MySejahtera & 1.0.19 & 
        \begin{tabular}[c]{@{}p{5cm}@{}}
            \cmark~Fixed the incorrect launch mode of an activity; \\ 
            \cmark~Removed three trackers: Google Analytics, Google CrashLytics, and Google Tag Manager.
        \end{tabular} \\ \midrule
        
        BlueZone & 2.0.2 & 
        \begin{tabular}[c]{@{}p{5cm}@{}}
            \cmark~Fixed potential privacy leakage. 
        \end{tabular} \\ 
        \midrule
        
        COVA Punjab & 1.3.11 &
        \begin{tabular}[c]{@{}p{5cm}@{}}
            \xmark~New WebView weakness is detected, which could enable popup windows; \\
            \xmark~New potential privacy leakage path found; \\
            \xmark~New tracker detected: Google CrashLytics and Google Ads.
        \end{tabular} \\ \midrule
        
        Coronavirus UY & 4.3.2 &
        \begin{tabular}[c]{@{}p{5cm}@{}}
            \xmark~New tracker detected: Google CrashLytics.
        \end{tabular} \\ \midrule
        
        Contact Tracer & N/A & 
        \begin{tabular}[c]{@{}p{5cm}@{}}
            \cmark~No longer available in Google Play Store 
        \end{tabular} \\ \bottomrule
    \end{tabular}
    \begin{tabular}{l}
       \cmark: Fixed, \xmark: New vulnerabilities found \\
    \end{tabular}
\end{table}

\subsection{Cases and Implications}\label{sec:security_case_study}

From our curated list of 40 apps, we select five typical apps around the world to further highlight key lessons we can learn with respect to security and privacy. The cases are TraceTogether, Next Step (DP3T), Private Kit, COVIDSafe, and Corona Warn App.  
\noindent\textbf{TraceTogether.~}

According to the \textsc{COVIDGuardian} analysis results, root detection~\cite{androidrootanditsproviders} has been implemented in TraceTogether. This potentially prevents SQL injection and data breaches, thereby reducing the risk to a certain extent. For example, in \texttt{o/C3271ax.java}, root detection logic is implemented by detecting the existence of specific root files in the system, \eg, \texttt{/system/app/Superuser.apk} and \texttt{/system/xbin/su}. By assessing their integrity, the app can detect whether a device is rooted and subsequently block users from either logging in or opening it. 

However, TraceTogether also includes a third-party customer feedback library, \textsc{zendesk SDK}, in which remote WebView debugging is enabled. This potentially allows attackers to dump the content in the WebView~\cite{devto2019dontleavedebugging}. When a user inputs confidential data, including passwords, in a debug-enabled WebView, attackers may be able to inspect all elements in the webpage~\cite{remotedebugwebview}. 
Fortunately, as per the static analysis, the only WebView with debugging mode enabled is to display articles; therefore, it does not contain confidential data.

\begin{mdframed}[backgroundcolor=black!10,rightline=false,leftline=false,topline=false,bottomline=false,roundcorner=2mm,everyline=true] 
	\textbf{Security guideline 1:} Never leave WebView with debugging mode enabled in the app release. 
\end{mdframed}
 \vspace{-2mm}

\noindent\textbf{Next Step (DP3T).~}
Next Step's database is not encrypted, and data is saved in plain text. In contrast to TraceTogether, the app does not implement any root detection capabilities. 
Although, the leakage of user's information via a rooted device may not be remotely exploitable, local malware may be able to gain root access. 
We therefore argue that root detection is necessary due to the large-scale deployment of such apps (over 60\% of the population~\cite{ferretti2020quantifying}), the long duration of their operation, and the fact that 7.6\% of Android users already have rooted their devices~\cite{rooted_device_worldwide}.
Further, in a rooted device, a malicious app could possibly access the database and manipulate COVID-19 contact records.

\begin{mdframed}[backgroundcolor=black!10,rightline=false,leftline=false,topline=false,bottomline=false,roundcorner=2mm,everyline=true] 
	\textbf{Security guideline 2:} To protect the database from being dumped and prevent data breaches, a solution should:
	\begin{enumerate}
	\item Implement database encryption~\cite{sqliteandroidencrypt}
	\item Enable root detection~\cite{rootdetection} and confidential data protection~\cite{supportdirectboot} at app startup.
	\end{enumerate}
\end{mdframed}
\vspace{-2mm}

In addition, as the database records timestamps and contact IDs, the leakage of the database from a rooted device could be exploited to mount linkage attacks by adversaries~\cite{linkageattack}. Therefore, if enough data in a region were collected, contact IDs and timestamps could be used to analyze movements~\cite{defendingagainstuseridentify}. 

\vspace{1mm} 

\noindent\textbf{Private Kit.~} \label{sec:security_case_study_safe_path}

Similar to Next Step (DP3T), Private Kit does not encrypt the database and contains plaintext data. Additionally, the app creates temporary JSON files to store users' location data. Without any encryption and root detection, the temporary JSON files could be dumped from a rooted device.

\begin{mdframed}[backgroundcolor=black!10,rightline=false,leftline=false,topline=false,bottomline=false,roundcorner=2mm,everyline=true] 
	\textbf{Security guideline 3:} To prevent potential data breaches, confidential data must not be stored in temporary files in plain text.
\end{mdframed}

\noindent\textbf{COVIDSafe.~}\label{sec:security_case_study_covidsafe}

COVIDSafe 1.0.11 stores all tracing histories (including contacted device IDs and timestamps) into an SQLite database using plain text. Since the app does not implement root detection logic, tracing histories may be leaked from root devices and therefore potential linkage attacks could be implemented~\cite{defendingagainstuseridentify}. However, in the latest version, COVIDSafe fixed this issue by encrypting the local database with a public key.

\vspace{1mm}
\noindent\textbf{Corona Warn App.}
We find two security features worth highlighting. Corona Warn App applies the \textsc{SQLCipher}~\cite{sqlcipher} framework, which enhances the \textsc{SQLite} database by making it more suitable for encrypted local data storage.  It also introduces the \textsc{conscrypt}~\cite{conscrypt} framework, which uses BoringSSL to provide cryptographic primitives and Transport Layer Security for Java applications on Android, during data transmission.  

\begin{mdframed}[backgroundcolor=black!10,rightline=false,leftline=false,topline=false,bottomline=false,roundcorner=2mm,everyline=true] 
	\textbf{Security guideline 4:} To protect local databases, introducing professional security frameworks is useful, \eg \textsc{conscrypt} and \textsc{SQLCipher} are open-source, well-documented, frequently-updated, widely-used, and their code is frequently reviewed. 
\end{mdframed}

\section {RQ3: Reviewing and Assessing Security and Privacy Threats}\label{sec:app_threats}

In this section, we review the major security and privacy threats facing contact tracing apps, based on our prior analysis. 

\subsection{User Privacy Exposure}
 We envisage three groups of contact tracing app users, based on their health status: 

\begin{itemize}[leftmargin=*]
    \item \textbf{Generic user.} A typical user of the contact tracing system, who is healthy or has not been diagnosed yet. 
    \item \textbf{At-risk user.} A user who has recently been in contact with an infected user. Ideally, an at-risk user will receive an at-risk alarm from the app.
    \item \textbf{Diagnosed user.} A diagnosed patient who will be asked to reveal their  information as well as the information of at-risk users to the health authorities, \eg, the diagnosis of their infection, their movement history, the persons they have been in contact with. 
\end{itemize}

\noindent Based on our prior analysis, we further define five broad categories of apps: 
\begin{itemize}[leftmargin=*]
    \item \textbf{Level I}~{\LOne}: ``No data is shared with a server or users'', the most secure level.
    \item \textbf{Level II}~{\LTwo}: ``Tokens are shared with proximity users'', a medium exposure level with only tokens containing no PII being exchanged between users. 
    \item \textbf{Level III}~{\LThree}: ``Tokens are shared with the server'', a medium exposure level with tokens exposed to the server.
    \item \textbf{Level IV}~{\LFour}: ``PII is shared with a server'', a high risk exposure level.
    \item \textbf{Level V}~{\LFive}: ``PII is released to public'', the highest risk exposure level. 
\end{itemize}

Based on the in-app instructions, documents provided in apps' official websites, and the privacy policies, we assess user privacy exposure and threats posed by the 40 apps listed in Table~\ref{table:app_list}.
We translated to English the materials not written in English with Google Translate. For 13 out of the 40 apps, we were not able to collect adequate information to conduct the privacy assessment. 
\textit{Significantly, 13 out of the 40 apps lacking transparent documentation were also identified as the ones with higher than the average number of security and privacy risks using \textsc{CovidGuardian}. Considering the lack of transparency of such apps with more than two million downloads, our findings demonstrate the pressing need for a holistic security and privacy assessment tool.}

As summarized in Table~\ref{table:privacy_exposure&attacks}, all 27 assessed apps have user privacy exposure to some extent. 
We determine user exposure level as IV in some centralized apps, such as COVIDSafe, TraceTogether, and apps from \#18 to \#27.
This is primarily because the central servers request users' PII (\eg, name, phone number, postcode, or even location information) during registration or execution. This could be fixed by better privacy by design, exemplified by \#4 StopCovid France, which does not collect such PII. 
Meanwhile, most decentralized Bluetooth-based apps are categorized as a lower-level exposure. This is because they utilize non-identifiable tokens, instead of directly using PII in contact tracing. However, we determine the privacy exposure level of diagnosed users in Hamagen as Level~V, as their location information could be accessed by third-party users, allowing the diagnosed users to be potentially re-identified. 

\subsection{Security and Privacy Threats}\label{sec:privacy_threats}

\begin{figure*}[t]
    \centering
    \includegraphics[width=1\linewidth]{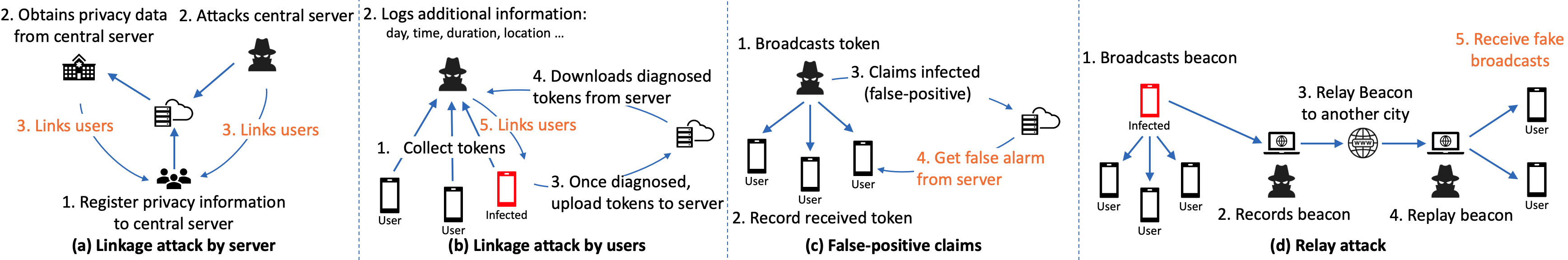}
    \vspace{-6mm}
    \caption{Four types of potential security and privacy threats: (a) \textbf{linkage attacks by a server:} in centralized systems, a key privacy concern is metadata leakage by the server; (b) \textbf{linkage attacks by users:} in systems based on information exchange between users, attackers could re-identify users using data published or received from users; (c) \textbf{false positive  claim:} attackers can incorrectly register as infected, which will generate false at-risk alerts; (d) \textbf{relay attacks:} in Bluetooth based apps, a malicious attacker can potentially redirect/replicate Bluetooth broadcasts from one place to another, thereby generating false at-risk alerts.}
    \label{fig:attacks}
    \vspace{-5mm}
\end{figure*}

To further examine security and privacy threats against contact tracing apps, we systematically search and review the state-of-the-art~\cite{chan2020pact,dp3tanalysis,baumgartner2020mind,gvili2020security}. 
We use Google Search, Google Scholar, and Twitter to discover research papers and Twitter comments or media outlets referring to  published or arXiv papers.
To achieve this, we use a set of associated search keywords (\eg, security and privacy attacks, COVID-19 contact tracing apps, vulnerabilities, effectiveness, safety). 
To expand this set, we further examine papers in the bibliographies of each keyword-filtered paper. 
We then manually exclude irrelevant topics and synthesize four dominant security and privacy threats:
\one~server link attacks; \two~user link attacks; \three~false positive claims; and \four~relay attacks. We summarize the nature of these threats in Figure~\ref{fig:attacks} and discuss them in turn below.

\begin{table}[t]
    \centering
    \caption{User privacy exposure and threats to apps.}
    \vspace{-3mm}
    \label{table:privacy_exposure&attacks}
    \resizebox{0.95\linewidth}{!}{
    \begin{tabular}{@{}r|l|lll|cccc|c@{}}
        \toprule
        \multirow{2}{*}{\textbf{\#}} &
        \multirow{2}{*}{\begin{tabular}[c]{@{}l@{}}\textbf{Apps}\\\\(27 of 40 apps assessed; \\13 apps do not provide \\adequate information)\end{tabular}} & 
        \multicolumn{3}{c|}{\textbf{User Privacy Exposure}} & \multicolumn{4}{c|}{\textbf{Threats}} & \multirow{2}{*}{\rotatebox[origin=c]{90}{\textbf{Architecture}}} \\ \cline{3-9}
        & & 
        \rotatebox[origin=c]{90}{\textbf{Generic}} & \rotatebox[origin=c]{90}{\textbf{At-risk}} & \rotatebox[origin=c]{90}{\textbf{Diagnosed}} & \rotatebox[origin=c]{90}{\textbf{Linkage-Server}} & \rotatebox[origin=c]{90}{\textbf{Linkage-User}} & \rotatebox[origin=c]{90}{\textbf{False-Claim}} & \rotatebox[origin=c]{90}{\textbf{Relay Attack}} & \\ \midrule
        1 & CovidSafe & \LTwo{} & \LFour{} & \LFour{} & \pie{360} & \pie{360} & \pie{90} & \pie{360} & C \\ 
        2 & HaMagen & \LOne{} & \LOne{} & \LFive{} & \pie{360} & \pie{360} & \pie{90} & \pie{90} & D\\
        3 & TraceTogether & \LTwo{} & \LFour{} & \LFour{} & \pie{360} & \pie{360} & \pie{90} & \pie{360} & C\\
        4 & StopCovid France & \LTwo{} & \LThree{} & \LThree{} & \pie{360} & \pie{360} & \pie{90} & \pie{360} & C \\
        5 & Next Step (DP3T) & \LTwo{} & \LThree{} & \LThree{} & \pie{90} & \pie{360} & (-) & \pie{360} & D \\
        6 & Corona Warn App & \LTwo{} & \LTwo{} & \LThree{} & \pie{90} & \pie{360} & \pie{90} & \pie{360} & D \\ 
        7 & NHS Test and Tracing App & \LTwo{} & \LTwo{} & \LThree{} & \pie{90} & \pie{360} & \pie{90} & \pie{360} & D \\
        8 & TraceCorona & \LTwo{} & \LTwo{} & \LThree{} & \pie{90} & \pie{360} & \pie{90} & \pie{360}  & D\\
        9 & Private Kit & \LTwo{} & \LTwo{} & \LThree{} & \pie{90} & \pie{360} & \pie{90} & \pie{360} & D\\
        10 & MySejahtera & \LFour{} & \LFour{} & \LFour{} & \pie{360} & \pie{90} & \pie{90} & \pie{90} & C \\
        11 & Smittestop &  \LTwo{} & \LThree{} & \LThree{} & \pie{90} & \pie{360} & \pie{90} & \pie{360} & C \\
        12 & COVID Alert & \LTwo{} & \LTwo{} & \LThree{} & \pie{90} & \pie{360} & \pie{90} & \pie{90} & D \\
        13 & SwissCovid & \LTwo{} & \LTwo{} & \LThree{} & \pie{90} & \pie{360} & \pie{90} & \pie{360} & D \\
        14 & Bluezone & \LTwo{} & \LTwo{} & \LThree{} & \pie{90} & \pie{360} & \pie{90} & \pie{360} & D \\
        15 & COCOA & \LTwo{} & \LTwo{} & \LThree{} & \pie{90} & \pie{360} & \pie{90} & \pie{360} & D\\ 
        16 & Immuni & \LTwo{} & \LTwo{} & \LThree{} & \pie{90} & \pie{360} & (-) & \pie{360} & D \\ 
        17 & Stopp Corona & \LTwo{} & \LTwo{} & \LThree{} & \pie{90} & \pie{360} & \pie{360} & \pie{360} & D \\
        18 & Aarogya Setu & \LFour{} & \LFour{} & \LFour{} & \pie{360} & \pie{360} & \pie{360} & \pie{360} & C \\
        19 & EHTERAZ & \LFour{} & \LFour{} & \LFour{} & \pie{360} & \pie{90} & \pie{90} & \pie{90} & C  \\
        20 & Vietnam Health Declaration & \LFour{} & \LFour{} & \LFour{} & \pie{360} & \pie{90} & \pie{360} & \pie{90} & C \\
        21 & STOP COVID19 CAT & \LOne{} & \LFour{} & \LFour{} & \pie{360} & \pie{90} & \pie{360} & \pie{90} & C \\
        22 & CG Covid-19 ePass & \LFour{} & \LFour{} & \LFour{} & \pie{360} & \pie{90} & \pie{90} & \pie{90} & C \\
        23 & StopTheSpread COVID-19 & \LOne{} & \LFour{} & \LFour{} & \pie{360} & \pie{90} & \pie{360} & \pie{90} & C \\
        24 & Stop COVID-19 KG & \LFour{} & \LFour{} & \LFour{} & \pie{360} & \pie{90} & (-) & \pie{90} & C \\
        25 & BeAware Bahrain & \LOne{} & \LFour{} & \LFour{} & \pie{360} & \pie{360} & \pie{360} & \pie{90} & C \\
        26 & Nepal COVID-19 Surveillance & \LOne{} & \LFour{} & \LFour{} & \pie{360} & \pie{90} & \pie{360} & \pie{90} & C \\
        27 & Stop Covid & \LTwo{} & \LTwo{} & \LFour{} & \pie{360} & \pie{360} & \pie{360} & \pie{360} & C \\
        \bottomrule
    \end{tabular}}
    \begin{tabular}[l]{@{}l@{}}
        \scriptsize{\pie{90}: the system is well protected~\pie{360}: the system is at-risk C: Centralized~D: Decentralized} \\\scriptsize{(-): Inadequate information to conduct an assessment.}
    \end{tabular}
    \vspace{-3mm}
\end{table}

\noindent \textbf{Linkage attacks by servers.} In centralized systems, the major privacy concern is metadata leakage by the server. 
Consequently, Grace will be able to collect a large amount of  PII, such as names, phone numbers, contact lists, post code, home addresses, location trails. Therefore Grace is able to deduce the social connections of Alice. 
However, in some decentralized systems (\eg, Hamagen), the server also learns users' PII, which can compromise such such decentralized systems too (as marked in Table~\ref{table:privacy_exposure&attacks}). 

\begin{mdframed}[backgroundcolor=black!10,rightline=false,leftline=false,topline=false,bottomline=false,roundcorner=2mm]
    \textbf{Privacy guideline 1:}~To protect users' privacy against linkage attacks by a server, a contact tracing app should: 
    \begin{enumerate}
    	\item Avoid sharing PII with central points or
    	\item Implement a decentralized design.
    \end{enumerate}
\end{mdframed}

\noindent \textbf{Linkage attacks by users.} 
Linkage attacks performed by users (Mallory), try to re-identify Alice or Bob. In contact tracing systems that directly publish users' PII, Bob is obviously at risk of privacy leakage. For other apps listed in Table~\ref{table:privacy_exposure&attacks} that rely on information exchange between users (\eg, DP3T, which implements an ephemeral ID design), Mallory is still able to identify Bob using more advanced attacks. For example, if Mallory places a Bluetooth receiver near Bob's home or working place and ensures that the device will only receive Bluetooth broadcasts from Bob. Once Bob is diagnosed, Mallory will receive an at-risk alarm and immediately acknowledge that the infected patient is Bob. 
In addition, Mallory can log the timestamp and the received ephemeral ID when in contact with Bob, which could be done by modifying the app or developing a customised app using the open-sourced contact tracing framework. 
Once Bob is diagnosed, Mallory will be able to trace back the source of recording and re-identify Bob and potentially infected users. 
Similar attacks were described as \textit{Paparazzi Attacks} and \textit{Nerd Attacks} in~\cite{dp3tanalysis}.
Even worse, if Mallory distributes multiple broadcast receivers in a large area they could potentially trace the movement of Bob by tracing the records on each device. 

\begin{mdframed}[backgroundcolor=black!10,rightline=false,leftline=false,topline=false,bottomline=false,roundcorner=2mm] 
	\textbf{Privacy guideline 2:} To protect users' privacy against linkage attacks by an adversary, a solution should: 
	\begin{enumerate}
	\item Avoid data sharing between users or
	\item Ensure privacy protections exist for any published data. 
	\end{enumerate}
\end{mdframed}

\noindent \textbf{False positive claims.} 
In some systems, Bob can register as infected via the app. If Mallory exploits such a mechanism and registers as a (fake) infected user, Alice will receive a false-positive at-risk alarm, which may cause social panic or negatively impact evidence-driven public health policies. 
Most solutions mitigate this issue by implementing an authorization process, \ie, a one-time-use permission code. 

\begin{mdframed}[backgroundcolor=black!10,rightline=false,leftline=false,topline=false,bottomline=false,roundcorner=2mm] 
    \textbf{Privacy guideline 3:} To protect a system against false-positive-claim attacks, a solution should establish an authorization process. 
\end{mdframed}

\noindent \textbf{Relay attacks.} To apply a relay attack, Mallory could collect existing broadcast messages exchanged between users, then replay them at another time, or forward them through proxy devices to a remote location. Due to the lack of message validation in solutions that utilize information broadcasts, a user will not be able to determine whether a received broadcast is from a valid source or from a malicious device. Thus, any received broadcast will be recorded as a contact event. A malicious attacker can potentially redirect all the traffic from one place to another, resulting in a targeted area being incorrectly locked-down.

\begin{mdframed}[backgroundcolor=black!10,rightline=false,leftline=false,topline=false,bottomline=false,roundcorner=2mm] 
    \textbf{Privacy guideline 4:} To protect a system against relay attacks, a solution should: 
    \begin{enumerate}
	    \item Either avoid utilizing information broadcast or 
	    \item Implement a validation approach.
	\end{enumerate}
\end{mdframed}

\noindent\textbf{Summary.}
The linkage attacks by a server are the overarching threats to centralized systems, as third-party attackers (or the server itself) are able to re-identify users (if PII is exposed to the central server). 
Although StopCovid France is robust to such a threat, it is still susceptible to re-identification risks since the information from the server enables linkage between anonymous IDs and the corresponding permanent app identifier. This permits the tracing of users based on IDs observed in the past, as well as future movements.

Additionally, apps that use Bluetooth broadcasts are exposed to linkage attacks by users.
We note that, although the false positive claims could be mitigated by authorization to allow only positive users to upload diagnosed data to the server, there are still some apps failing to implement essential authorization. Furthermore, the relay attack is another threat causing false positives in Bluetooth-based apps. Notably, although TraceCORONA is rated as at-risk against relay attacks, its specific design provides protection against one-way-relaying unlike other apps.

\begin{mdframed}[backgroundcolor=white!10,roundcorner=2mm,everyline=true] 
	\textbf{Answer to RQ3.} 
	We manually rate the level of user privacy exposure for each contact tracing app. We further categorize the robustness of apps against security and privacy threats. 
	We ascertain that the apps in question are vulnerable to at least one of the four threats. Not all decentralized architectures are necessarily more secure than those adopting centralized architectures, \eg, Hamagen, a decentralized location-based system. 
\end{mdframed}

\section{RQ4: User Study}\label{sec:user_study}

In this section, we present a survey, exploring the user perceptions of contact tracing apps. Our objective is to query the \textit{likelihood} and \textit{concerns} associated with using the apps.
Through the study we aim to ascertain user concerns and requirements of contact tracing apps (\textit{RQ4}).

\begin{table*}[t]
	\centering
	\vspace{2mm}
	\begin{minipage}[t]{.47\linewidth}
	    \includegraphics[width=1\linewidth]{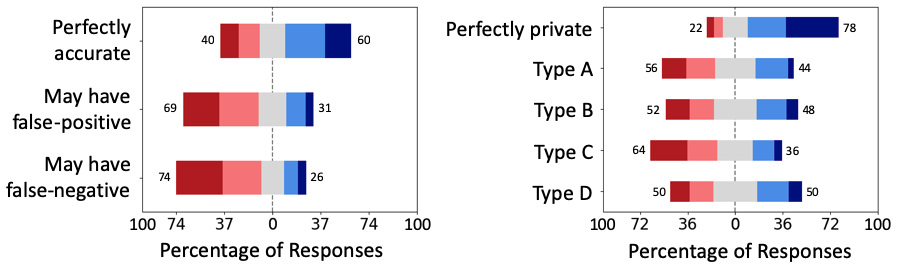}
		\vspace{-6mm}
		\captionof{figure}{(a)~Participants' likelihood of using contact tracing apps \vs the accuracy of proximity detection; (b)~Participants' likelihood of using apps across different privacy-preserving and data sharing scenarios.  \firebrick~: 5 = extremely unlikely, \darkblue~: 1 = extremely likely.}
		\label{fig:likelihood}
    \end{minipage}\quad
    \begin{minipage}[t]{.225\linewidth}
    	\includegraphics[width=1\linewidth]{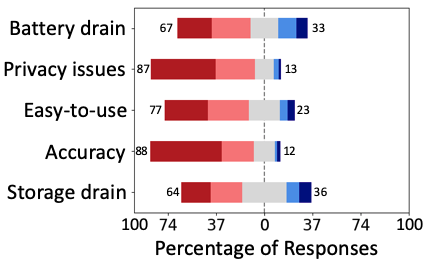}
    	\vspace{-6mm}
    	\captionof{figure}{Participants' concerns about contact tracing apps. \firebrick~: 5 = extremely concerned, \darkblue~: 1 = extremely unconcerned.}
    	\label{fig:concerns}
    \end{minipage}\quad
    \begin{minipage}[ht]{.25\linewidth}
        \vspace{-4mm}
        \captionof{table}{Mann-Whitney results on the likelihood of using contact tracing apps.}
        \includegraphics[width=1\linewidth]{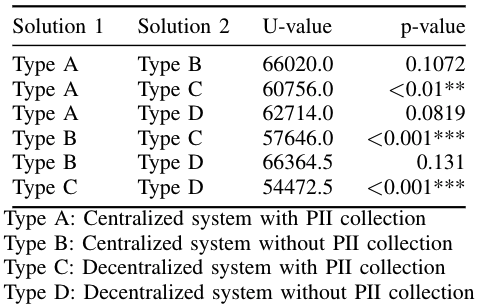}
    	\label{table:solution_comparison}
	\end{minipage}
	\vspace{-8mm}
\end{table*}

\subsection{Participant Recruitment}

The proportion of youth aged 15-24 years with COVID-19 has increased six-fold from 24 February through 12 July 2020~\cite{who_report198}. It was reported by Reuters~\cite{reuters} that young people who are visiting nightclubs and beaches are causing a rise in fresh cases with potential consequences for more vulnerable age groups. Therefore, we focus our user study on 18-29 year olds. 
We have recruited 373 volunteers and asked them standard demographic questions, \ie, age range, education, gender, and nationality. 
All participants reside in Australia but have various cultural backgrounds (58\% from Oceania, 20\% from Asia, and 14\% from other geographic regions; 30 participants did not indicate their nationalities). 39\% of participants are male (59\% female). One participant identified themselves as other gender and six participants refrained. 67\% of participants are university graduates and 30\% are high school graduates.

\subsection{Survey Protocol}\label{sec:survey-protocol}
 
To calibrate the scope of our survey without introducing bias, we provide vignettes describing user privacy exposure levels using the security and privacy threats reviewed in Section~\ref{sec:app_threats}. 

All questions, excluding demographics, are five-point Likert scale~\cite{likert1932technique} questions. The Likert scale is a quantitative, fine-grained, and user-friendly method of collecting data from users. The scales ask participants to indicate how much they are likely (1) or unlikely (5), to be unconcerned (1) or concerned (5) about using contact tracing apps. The survey adopts a bespoke model and integrates questions from related survey instruments used by Simko~\etal~\cite{simko2020covid} and Kaptchuk~\etal ~\cite{kaptchuk2020good}. 
We provide the survey design and questionnaire in our open source website and summarize the two item categories below. 

\noindent\textbf{(a) Likelihood of using contact tracing apps.}
To investigate the usability requirements of contact tracing apps, we asked participants their likelihood of using contact tracing apps under different scenarios: \one~functionality scenarios, \ie, if the accuracy of proximity contact detection and at-risk alarm generation is not perfect (false positives or false negatives exist); and \two~privacy scenarios, \ie, if personal data (location, phone number, postcode, or anonymous identifier) will be shared with other entities (other users or governments), or if the authorities require them to provide data (location, proximity data, or anonymous tokens) after being diagnosed.

Notably, \textit{to ensure the survey is designed for both users and non-users, we did not explicitly mention any specific contact tracing app}, but still covered all contextual privacy concepts of state-of-the-practice contact tracing apps discussed in Section~\ref{sec:solutions}. For example, COVIDSafe is covered by scenarios in which a ``phone number and postcode are shared with governments or health authorities'' and if the app will ``upload proximity data if tested positive''. 
Similarly, the Corona Warn App or others that implement the Google and Apple (Gapple) framework are covered by scenarios where ``anonymous identifiers are shared with other users'' and the app will ``upload anonymous tokens if being tested positive''.

\noindent\textbf{(b) Concerns about use of contact tracing apps.} We also asked the participants about their concerns regarding contact tracing apps. We focused mainly on three aspects: \one~the usability of contact tracing apps, including battery drain, storage drain, and ease-of-use; \two~the effectiveness of contact tracing apps, \ie, to what extent users are concerned about the accuracy of contact tracing; and \three~the concerns about privacy. 
We conducted the survey through both pencil-and-paper and SoGoSurvey~\cite{sogosurvey}. 3.4\% (15) of participants used the paper version of the survey and there is no significant impact on the study results. During the paper survey, we did not receive any queries from participants.

\subsection{Data Analysis and Results}

Our variables are ordinal and the responses to each question are not expected to be normally distributed. Therefore, we use a Mann-Whitney U-test~\cite{mann1947test} for statistical significance testing to ascertain the key factors that impact user concerns and requirements (or expectations).  
Concretely, we aggregate item responses to assess the participants' likelihood of using contact tracing apps across \textit{four} different  privacy-preserving  and  data  sharing  scenarios (as described in the Section~\ref{sec:privacy_threats}).

We calculate the $U$-value and $p$-value of each pair of contact tracing solutions among centralized solutions with PII collected (Type~A), centralized solutions with non-PII collected (Type~B), decentralized solutions with PII collected (Type~C), and decentralized solutions with non-PII collected (Type~D).  If the $p$-value is smaller than 0.05, we reject the null hypothesis that users are equally likely to use any of the two contact tracing apps. We use \textsc{SciPy.stats} for our analysis.
 
\noindent\textbf{(a) The accuracy of contact tracing impacts the likelihood of app use. Users are sensitive to sharing PII in a decentralized system.} 
As shown in Figure~\ref{fig:likelihood}(a), if a contact tracing app can accurately detect proximity and notify users who may be at-risk, more than 60\% of participants are likely to use it. However, in reality, it is hard to eliminate false-positives and false-negatives. When the tracing accuracy concern is considered, the proportion of users likely to use contact tracing apps drops to 31\% and 26\%, respectively. Our results align well with a survey of COVID-19 app usage taken in the USA~\cite{kaptchuk2020good}.

\begin{comment}
\begin{table}[t]
    \centering
    \caption{Mann-Whitney results on the likelihood of using contact tracing apps}
    \label{table:solution_comp}
    \vspace{-2mm}
    \begin{tabular}{@{}llrr@{}}
        \toprule
        Solution 1      & Solution 2 & \multicolumn{1}{c}{$U$-value} &
        \multicolumn{1}{c}{$p$-value} \\ \midrule
        Type A     & Type B     & 66020.0 & 0.1072  \\
        Type A     & Type C     & 60756.0 & \textless 0.01**    \\
        Type A     & Type D     & 62714.0 & 0.0819  \\
        Type B     & Type C     & 57646.0 & \textless 0.001*** \\
        Type B     & Type D     & 66364.5 & 0.131   \\
        Type C     & Type D     & 54472.5 & \textless 0.001***    
\\ \bottomrule
    \end{tabular}
    \begin{tabular}[c]{@{}l@{}}
        Type A: Centralized system with PII collection \\
        Type B: Centralized system without PII collection \\
        Type C: Decentralized system with PII collection \\
        Type D: Decentralized system without PII collection\end{tabular}
\end{table}
\end{comment}

Figure~\ref{fig:likelihood}(b), shows how likely users are to use a contact tracing app in different scenarios. 
Here, 50\% of participants listed positive responses to decentralized apps with non-PII identification collected and shared (Type~D). This resonates with the fact that Bluetooth-based decentralized systems preserve user privacy more than centralized systems. However, decentralized apps with PII collected (Type~C) are not as popular with users. More than 64\% of participants said that they are unlikely to use such an app. 
It can be seen that even this young cohort of users is sensitive to sharing PII. 

Table~\ref{table:solution_comparison} presents the Mann-Whitney U-tests comparing pairs of contact tracing apps. 
Our results indicate that the differences between the likelihood of using decentralized apps with PII collected (Type~C) and other apps is statistically significant.  
This is perhaps due to the privacy design in a decentralized PII-collected system. That is, diagnosed users' information (\eg, location information) will be collected and shared with other users, while the other types of apps do not share such information between users.
When compared with the other three types, we had $p$-values greater than 0.05. Hence, despite the different levels of privacy protection, there is no statistically significant difference in the likelihood of users using these apps. However, from Figure~\ref{fig:likelihood}(b), we can infer that it is more likely for users to accept and use the decentralized apps without PII collection (Type D).

Additionally, as evidenced in Figure~\ref{fig:likelihood}, we find that more than 78\% of participants are likely to use perfectly private contact tracing apps. This indicates a much higher likelihood than those who prefer a perfectly accurate contact tracing app (60\%). The difference is statistically significant ($p\textrm{-value} < 0.0001$), indicating that users are more likely to accept and use contact tracing apps that satisfy privacy protection requirements.

\noindent\textbf{(b) User concerns focus on privacy and tracing accuracy.} 
As shown in Figure~\ref{fig:concerns}, more than 55\% of participants are extremely concerned about the tracing accuracy of apps, and more than 49\% of participants are extremely concerned about privacy issues. 

\subsection{Threats to Validity of Our User Study}
\noindent \textbf{External validity.}
The participants of the survey were all from one country, and the duration of the survey was three weeks. 
Further, our survey may be subject to volunteer bias and non-response bias~\cite{brassey2017volunteer} (\ie, participants and their responses may have different characteristics from the general population of interest).
To mitigate the volunteer bias, we designed the survey questionnaire to be completed in 10-15 minutes and anonymized the responses. 

\noindent\textbf{Internal validity.}
Considering the nature of the user study, participants may find that some questions are confusing or unclear. This will lead to incorrect or inconsistent responses. To mitigate this threat, we conducted a pilot survey with 45 computer science students and updated the questions based on feedback. In addition, as the responses are anonymous, there could be participants who took the survey multiple times.

\begin{mdframed}[backgroundcolor=white!10,rightline=true,leftline=true,topline=true,bottomline=true,roundcorner=2mm,everyline=true] 
 	\textbf{Answer to RQ4.} Privacy design and tracing accuracy impact the likelihood of app use. Furthermore, compared to users' expectations of tracing accuracy, users are more likely to accept and use apps with better privacy by design. Interestingly, if PII data is collected, users prefer a centralized solution in contrast to a decentralized solution that collects PII data.
\end{mdframed}

\vspace{-1mm}

\section{Conclusion}\label{sec:lessons_learned}

This study has developed a security and privacy assessment tool, \textsc{COVIDGuardian}. This tool can evaluate the security weaknesses, vulnerabilities, potential privacy leaks, and malware in contact tracing apps. Using \textsc{COVIDGuardian}, we have conducted a comprehensive empirical security and privacy assessment of 40 contact tracing apps. Our results have identified multiple security and privacy risks, as well as threats. 
Naturally, our analysis has confirmed that no apps can protect users' security and privacy against \emph{all} potential threats.
To understand the perception of users, we have also performed a survey involving 373 participants. This has further consolidated our observations of user concerns. In the future, we plan to extend our study to obtain user feedback from a wider geographic and demographic range. Examining network traffic originating from contact tracing apps is also worth further exploration. 

\section*{Acknowledgments}
We would like to thank the reviewers for their valuable feedback. This work was supported in part by the Australian Research Council (ARC) Discovery Project (DP210102670) and the EPSRC (EP/S033564/1).

\newpage
\bibliographystyle{IEEEtranS}

{\small\selectfont\bibliography{reference}}

\end{document}